\documentclass{PoS}

\title{
\rule{0cm}{2.5cm}\vspace{-4.5cm}\\{
\hfill \it \normalsize SB/F/377-10
 } \vspace{2cm}\\Associated production of a $W$ or $Z$ boson with bottom
quarks at the Tevatron and the LHC}

\ShortTitle{Associated production of a $W$ or $Z$ boson with bottom quarks}

\author{Fernando Febres Cordero\\
        Universidad Sim\'on Bol\'{\i}var, Departamento de F\'{\i}sica, 
        Apartado 89000, Caracas 1080A, Venezuela \\
        E-mail: \email{ffebres@usb.ve}}

\author{Laura Reina\\
        Florida State University, Physics Department, Tallahassee, FL 32306-4350, USA\\
        E-mail: \email{reina@hep.fsu.edu}}

\author{\speaker{Doreen Wackeroth}\\
        University at Buffalo, The State University of New York, Department of Physics, Buffalo, NY 14260-1500, USA\\
        E-mail: \email{dow@ubpheno.physics.buffalo.edu}}

\abstract{ 

 We present total and differential cross sections for $Wb\bar{b}$ and
 $Zb\bar{b}$ production at the CERN Large Hadron Collider including
 Next-to-Leading Order (NLO) QCD corrections and full bottom-quark
 mass effects.  We discuss the scale uncertainty of the total cross
 sections due to the residual renormalization- and factorization-scale
 dependence of the truncated perturbative series.  We also discuss
 $b$-quark mass effects in kinematic distributions by comparing with a
 calculation that considers massless bottom quarks, as implemented in
 the Monte Carlo program MCFM. The effects of a non-zero bottom-quark
 mass ($m_b$) cannot be neglected in phase-space regions where the
 relevant kinematic observable, such as the transverse momentum of the
 bottom quarks or the invariant mass of the bottom-quark pair, are of
 the order of $m_b$.  Finally, we present the result of a detailed
 comparison of NLO QCD predictions for $W+b$-jet production with one
 or two jets with Tevatron data.

}

\FullConference{RADCOR 2009 - 9th International Symposium on Radiative Corrections (Applications of Quantum Field Theory to Phenomenology) ,\\
		 October 25 - 30 2009\\
		 Ascona, Switzerland}

\begin{document}

\section{Introduction}

The associated production of a weak gauge boson with one or two
$b$ jets constitutes not only an important background process to Higgs
boson searches for light Higgs bosons ($M_H \ {\stackrel{<}{\scriptstyle \sim}} \ 135$~GeV), to
single top-quark production and many searches for signals of new
physics, but also represents a unique opportunity to test and improve
the theoretical prediction for heavy quark jets at hadron
colliders. The cross sections for $W$ and $Z$ boson production with
$b$ jets has been measured at the Tevatron $p \bar p$ collider at
Fermilab ($\sqrt{s}=1.96$ TeV) by both the
CDF~\cite{Aaltonen:2008mt,Aaltonen:2009qi} and D0~\cite{d0}
collaborations, and these measurements will continue to improve once
more data have been analyzed.  Studying the same cross sections in the
very different kinematic regimes available at the LHC $pp$ collider
will then be of great interest and will represent a crucial test of
our understanding of QCD at high-energy colliders.

The production of a $W$ or a $Z$ boson with up to two jets, one of
which is a $b$ jet, has been calculated including NLO QCD corrections
in the variable-flavor scheme (VFS) or five-flavor-number scheme
(5FNS)~\cite{5fns}, while the production of a $W$ or $Z$ boson with
two $b$ jets has been derived at NLO in QCD using the fixed-flavor
scheme (FFS) or four-flavor-number scheme (4FNS), first in the
massless $b$-quark approximation~\cite{4fnsmassless,Campbell:2003hd}
and more recently including full $b$-quark mass
effects~\cite{FebresCordero:2006sj,FebresCordero:2008ci,Cordero:2008ce,Cordero:2009kv}.
In the FFS or 4FNS only massless-quark densities are considered in the
initial state, while in a VFS or 5FNS an initial-state $b$-quark
density is introduced. The two schemes amount to a different ordering
of the perturbative series for the production cross section: in the
4FNS the perturbative series is ordered strictly by powers of the
strong coupling $\alpha_s$, whereas in the 5FNS the introduction of a
$b$-quark parton distribution function (PDF) allows to resum terms of
the form $\alpha_s^n \ln(m_b^2/M^2)^m$ at all orders in $\alpha_s$
(for fixed order of logarithms $m$), where $M$ represents the upper
integration limit of the $b$-quark transverse momentum and can be
thought to be of the order of $M_W$ or $M_Z$.  While the two
approaches can give very different results at leading order (LO) in
QCD, starting at NLO in QCD the total cross sections have been shown
to be consistent within their respective theoretical uncertainties for
both $H+1\,b$-jet production~\cite{h1bjet} (for a brief review see
also Ref.~\cite{Dawson:2004wq}) and single-top
production~\cite{Campbell:2009ss}.  In kinematical distributions,
however, $b$-quark mass effects which are fully taken into account in
the 4FNS can have a significant impact, particular in phase space
regions where the relevant kinematic observable is of the order of
$m_b$. Bottom-quark mass effects are also important in $W+1\,b$-jet
production and improved predictions at NLO QCD are provided
in~\cite{Campbell:2008hh}, where the calculations of $Wb\bar b$
production in the 4FNS with massive $b$ quarks and $Wbj$ production in
the 5FNS have been merged.  A similar study is currently in progress
for $Z+1\,b$-jet production~\cite{Zb_inprep}.  Improving the
predictions for $Z+1\,b$-jet production will be particularly relevant
at the LHC, where this process allows for a direct determination of
the $b$-quark PDF, to be used in the prediction of $H+1\,b$-jet
production, a discovery channel for beyond-the-SM Higgs bosons with
enhanced $b$-quark Yukawa couplings.

In these proceedings, we present results for $Wb\bar{b}$ and
$Zb\bar{b}$ production at the LHC, i.e. with both $b$ jets tagged in
the final state, keeping the $W$ and $Z$ boson on shell. We discuss
the impact of the $b$-quark mass on total kinematic distributions by
comparing with a calculation which neglects the $b$-quark mass at NLO
in QCD, as implemented in the MCFM package~\cite{MCFM:2004}.  Finally,
we present the results of a detailed comparison of NLO QCD predictions
for $W+b$-jet production with one or two jets with Tevatron data.

\section{$Wb\bar b$ and $Zb \bar b$ production at NLO QCD}

At tree level, the production of a $W$ boson with a pair of bottom quarks
consists of just one process, $q\bar{q}^\prime\rightarrow Wb\bar{b}$, while
$Zb\bar b$ production consists of two channels, namely $q\bar{q}\rightarrow
Zb\bar{b}$ and $gg\rightarrow Zb\bar{b}$.  In order to compute these processes at
NLO in QCD one needs to include one-loop virtual corrections to the tree-level
processes, as well as all real radiation corrections with up to one extra
parton in the final state, i.~e.  $q\bar{q}^\prime\rightarrow Wb\bar{b}+g$,
$gg,q\bar q\rightarrow Zb\bar{b}+g$, and the $qg$-initiated processes,
$qg(\bar{q}g)\rightarrow Wb\bar{b}+q^\prime(\bar{q}^\prime)$ and
$qg(\bar{q}g)\rightarrow Zb\bar{b}+q(\bar{q})$. Details of the calculations
and numerical results are provided in
Refs.~\cite{FebresCordero:2006sj,FebresCordero:2008ci,Cordero:2008ce,Cordero:2009kv}.
 
The LO results are based on the one-loop evolution of $\alpha_s$ and
the CTEQ6L1 set of PDFs~\cite{Lai:1999wy}, with
$\alpha_s^{LO}(M_Z)=0.130$, while the NLO results use the two-loop
evolution of $\alpha_s$ and the CTEQ6M set of PDFs, with
$\alpha_s^{NLO}(M_Z)=0.118$. We use the $k_T$ jet algorithm with a
pseudo-cone size of $R=0.7$ and we checked that our implementation of
the $k_T$ jet algorithm coincides with the one in MCFM.  We require
all events to have a $b$-jet pair in the final state, with a
transverse momentum either larger than $15$~GeV or $25$~GeV and
require the pseudorapidities of both $b$ jets satisfy
$|\eta^{b,\bar{b}}|<2.5$. We impose the same $p_T$ and $|\eta|$ cuts
also on the extra jet that may arise due to hard non-collinear real
emission of a parton, i.~e. in the processes $W/Zb\bar{b}+g$ or
$W/Zb\bar{b}+q(\bar{q})$. This hard non-collinear extra parton is
treated either \emph{inclusively} or \emph{exclusively}. In the
\emph{inclusive} case we include both two- and three-jet events, while
in the \emph{exclusive} case we require exactly two jets in the
event. Two-jet events consist of a $b$-jet pair that may also include
a final-state light parton (gluon or quark) due to the applied
recombination procedure. On the other hand, three-jet events consist
of events containing a $b$-jet pair plus an extra light jet. We notice
that, at NLO in QCD, all jets in three-jet events consist of a single
parton.

\begin{figure}[t]
\begin{center}
\includegraphics*[clip,scale=0.46]{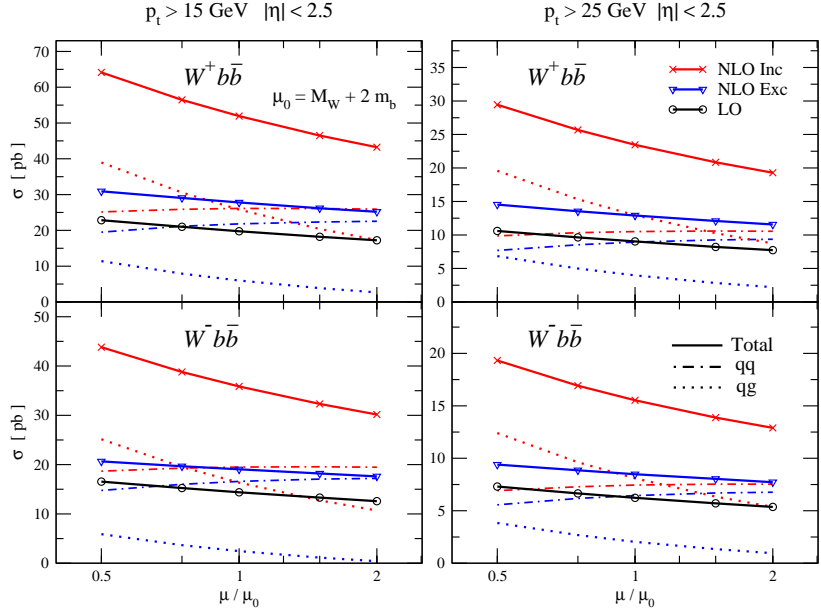} 
\caption[]{Dependence of the LO (black, solid), NLO \emph{exclusive}
  (blue, solid), and NLO \emph{inclusive} (red, solid) total cross
  sections for $W^+b\bar{b}$ and $W^-b\bar{b}$ production on the
  renormalization/factorization scales at the LHC ($\sqrt{s}=14$~TeV), including full $b$-quark mass
  effects, when $\mu=\mu_r=\mu_f$ is varied between $\mu_0/2$ and
  $2\mu_0$ (with $\mu_0=M_W+2m_b$). We also show the individual parton-level
  channels, $q\bar{q}^\prime$ (dash-dotted) and $qg+\bar{q}g$
  (dotted), for the \emph{inclusive} (red) and \emph{exclusive}
  (blue) cases. Taken from Ref.~\cite{Cordero:2009kv}.}
\label{fig:Wbb_mudep}
\end{center}
\end{figure}

\begin{figure}[htb]
\begin{center}
\includegraphics*[clip,scale=0.46]{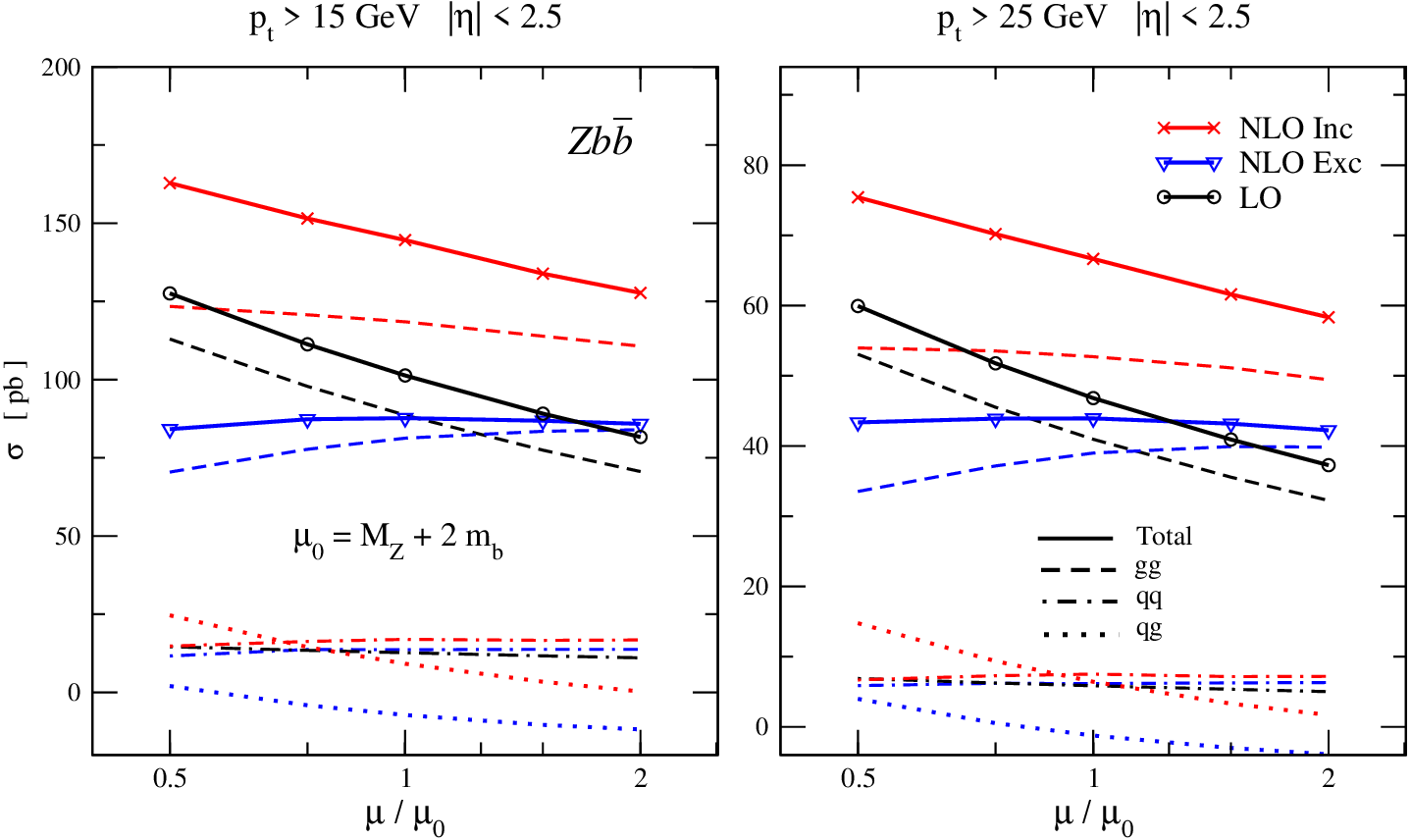} 
\caption[]{Dependence of the LO (black, solid), NLO \emph{exclusive}
  (blue, solid), and NLO \emph{inclusive} (red, solid) total cross
  sections for $Zb\bar{b}$ production on the
  renormalization/factorization scales at the LHC ($\sqrt{s}=14$~TeV), including full $b$-quark mass
  effects, when $\mu=\mu_r=\mu_f$ is varied between $\mu_0/2$ and
  $2\mu_0$ (with $\mu_0=M_Z+2m_b$). We also show the individual parton-level
  channels, $q\bar{q}^\prime$ (dash-dotted), $qg+\bar{q}g$ (dotted)
  and $gg$ (dashed), for LO and NLO (\emph{inclusive} (red) and
  \emph{exclusive} (blue)). Taken from Ref.~\cite{Cordero:2009kv}.}
\label{fig:Zbb_mudep}
\end{center}
\end{figure}

In Fig.~\ref{fig:Wbb_mudep} ($W^\pm b\bar b$) and
Fig.~\ref{fig:Zbb_mudep} ($Zb\bar b$) we illustrate the
renormalization- and factorization-scale dependence of the LO and NLO
total cross sections obtained for a massive $b$ quark, when
$\mu=\mu_r=\mu_f$ is varied between $\mu_0/2$ and $2\mu_0$, with
$\mu_0=M_V+2m_b$ ($V=W,Z$). We immediately notice that the impact of
NLO QCD corrections is very large, in particular in the
\emph{inclusive} case in $Wb\bar b$ production, where they increase
the LO cross section by a factor between two and three depending on
the scale. We also notice that the scale dependence of the NLO $Wb\bar
b$ cross section is worse than (\emph{inclusive} case) or comparable
to (\emph{exclusive} case) the scale dependence of the LO cross
section. This is different from what has been observed for the
Tevatron~\cite{FebresCordero:2006sj,Cordero:2008ce}, and was first
pointed out in a calculation with massless bottom
quarks~\cite{Campbell:2003hd}.  It is just a reminder of the fact
that, at a given perturbative order, the uncertainty due to the
residual renormalization- and factorization-scale dependence may
underestimate the theoretical uncertainty due to missing higher-order
corrections. A realistic determination of this uncertainty is usually
much more complex and requires a thorough understanding of the
perturbative structure of the cross section, in particular at the
lowest orders of the perturbative expansion.  In both $Wb\bar b$ and
$Zb \bar b$ production the NLO QCD corrections introduce a new
production channel not present at LO, the $qg$-initiated processes,
and as such introduces a LO scale dependence.  The NLO $Wb\bar{b}$ total cross section is particularly
affected by this process because there is no $gg$-initiated process at
LO.  In $Zb\bar b$
production, however, this process is not dominant at NLO and its impact is therefore less
pronounced. Indeed, the scale dependence of the \emph{exclusive}
$Zb\bar b$ cross section actually greatly improves at NLO in QCD,
while the scale dependence of the \emph{inclusive} one is only mildly
better than at LO, but not worse as it is the case in $Wb\bar b$
production.   Finally, the impact of the $qg$-initiated processes on the scale
dependence of the total cross sections is larger in the
\emph{inclusive} than in the \emph{exclusive} case because the
\emph{exclusive} cross section by definition discriminates against
processes with more than two jets in the final state.

\begin{figure}[htb]
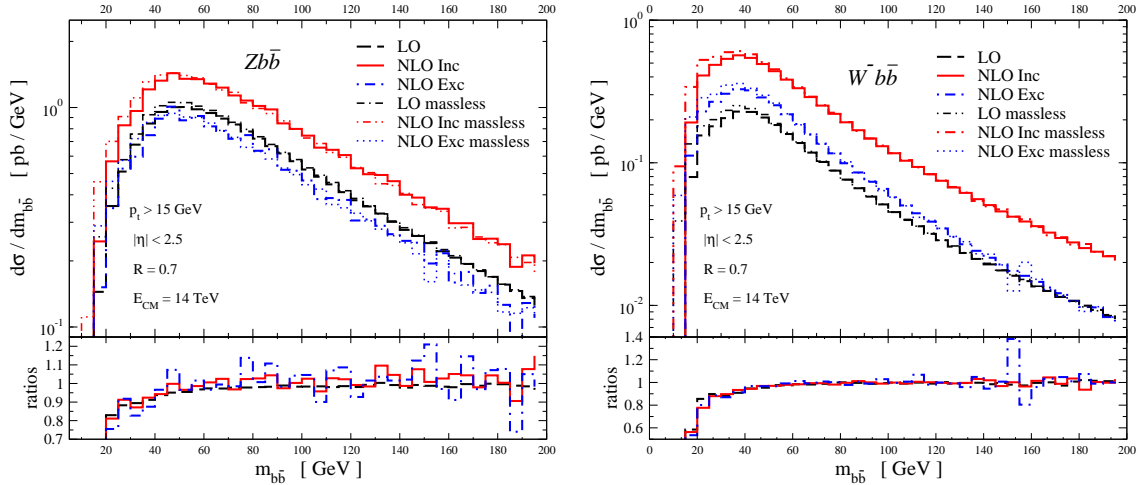

\begin{center}
\begin{tabular}{ll}
\includegraphics*[clip,scale=0.42]{Zbb_LHC_15_comp_MCFM_mbb} &
\includegraphics*[clip,scale=0.42]{Wmbb_LHC_15_comp_MCFM_mbb} 
\end{tabular} 
\caption[]{LO (black), NLO \emph{inclusive} (red) and NLO
  \emph{exclusive} (blue) $m_{b\bar{b}}$ distributions for Zbb (left hand side) and
  $Wbb$ (right hand side) 
  production at the LHC derived with $m_b\ne 0$ (LO: dashed,
  NLO \emph{inclusive}: solid, NLO \emph{exclusive}: dash-dotted) and
  from MCFM with $m_b=0$ (LO: double-dashed/dotted, NLO
  \emph{inclusive}: dashed/double-dotted, NLO \emph{exclusive}:
  dotted).  The lower window shows the ratio of the distributions for
  massive and massless $b$ quarks, $d\sigma(m_b\neq 0)/d\sigma(m_b=0)$
  (LO: dashed, NLO \emph{inclusive}: solid, NLO \emph{exclusive}:
  dash-dotted). Taken from Ref.~\cite{Cordero:2009kv}.}
\label{fig:mbb_comp_pt15}
\end{center}
\end{figure}

In Fig.~\ref{fig:mbb_comp_pt15}, we illustrate $b$-quark mass effects
on the example of $m_{b\bar b}$ distributions in $W^-b\bar b$ and
$Zb\bar b$ production.  These effects can impact the shape of the
kinematic distributions in particular in phase space regions where the
relevant kinematic observable is of the order of $m_b$. These effects
can be approximated by rescaling the NLO cross section for $m_b=0$
with the ratio of LO cross sections for massive and massless bottom
quarks as discussed in detail
in~\cite{FebresCordero:2006sj,FebresCordero:2008ci}. The total
production cross sections are reduced by $b$-quark mass effects, and
the effect is more pronounced the smaller the applied $p_T^b$
cut. However, these effects are in most cases smaller than the
residual scale dependence at NLO in QCD, especially in $Wb\bar b$
production for the \emph{inclusive} case.

\section{Comparison of $W+b$-jet predictions with Tevatron data} 

A recent measurement of the cross section of $W$ boson production in
association with one or two $b$ jets by the CDF collaboration at the Tevatron
finds~\cite{Aaltonen:2009qi}
\[\sigma_{b-\mathrm{jets}}\times \mathcal{B} (W\rightarrow \ell \nu ) (\mathrm{CDF})
= 2.74 \pm 0.27\textrm{(stat.)} \pm 0.42\textrm{(syst.)}~\mathrm{pb}\] 
This $b$-jet cross section includes $W+2 b$-jets and $W+1 b$-jet contributions,
and the NLO QCD predictions for both signatures have to be considered. The NLO
QCD prediction for $W+2 b$-jets production is based on
Refs.~\cite{FebresCordero:2006sj,Cordero:2008ce} and the one for $W+1 b$-jet
production on Ref.~\cite{Campbell:2008hh}, where in both cases events
with a non-$b$ jet that result in a three-jet event are discarded. In
Ref.~\cite{Campbell:2008hh}, the calculations of $Wb\bar b$ production in the
4FNS and $Wbj$ production in the 5FNS have been merged by consistently
treating contributions which arise in both processes. In this way, the
$b$-quark mass is taken into account, which cannot be neglected when one $b$
quark is treated fully inclusively, the tree-level $qg$-initiated process
discussed earlier is improved by including NLO QCD corrections in the 5FNS,
and large initial-state logarithms are resummed via the $b$-quark PDF
approach. As a result of this improved treatment, theoretically more stable
predictions for $W+1 b$-jet production are now available. Finally, combining
the results of predictions for $W+2 b$-jets and $W+1 b$-jet production, thereby
doubling the weights of $2 b$-jet events to obtain jet-level cross sections,
yields the following NLO QCD prediction (with
$\mu_r=\mu_f=M_W$ for the central value)~\cite{jfl}:
\[\sigma_{b-\mathrm{jets}}\times \mathcal{B} (W\rightarrow \ell \nu ) (\mathrm{NLO\; QCD})
= 1.22 \pm 0.14\textrm{(scale uncert.)}~\mathrm{pb} \; .\]
Together with the LO prediction of $0.91^{+0.29}_{-0.20}$~pb (including scale
uncertainties) this results in a moderate K-factor of about 1.34.  The
origin of the discrepancy between theory and experiment is presently under
investigation~\footnote{See also talks given at the workshop {\em Northwest
    Terascale Research Projects $W + b$ quark physics at the LHC}, held at the
  University of Oregon, http://physics.uoregon.edu/$\sim$soper/TeraWWW2.}.

\acknowledgments{

We would like to thank the organizers of the workshop {\em Northwest
Terascale Research Projects $W + b$ quark physics at the LHC},
especially Davison Soper, for the kind hospitality extended to us during
our visit at the University of Oregon, where part of this work was
discussed.  F.~F.~C. and L.~R. are also grateful to John Campbell,
Christopher Neu, and Evelyn Thomson for their collaboration on the
comparison of $W+b$-jets cross section with CDF data.  The work of
L.~R.~is supported in part by the U.S. Department of Energy under
grant DE-FG02-97IR41022.  The work of D.~W.~is supported in part by
the National Science Foundation under grants NSF-PHY-0547564 and
NSF-PHY-0757691, and by a DFG Mercator Visiting Professorship during
D.W.'s sabbatical leave at the Karlsruhe Institute of Technology
(KIT).

}

\end{document}